\documentclass[12pt]{article}
\usepackage{epsf,epsfig,amssymb,amsmath,graphicx,float}
\usepackage{color}

\begin{document}
\renewcommand{\thefootnote}{\fnsymbol{footnote}}

\begin{titlepage}

\begin{center}

\vspace{1cm}

{\Large {\bf Abundance of Asymmetric Dark Matter in Brane
    World Cosmology }}

\vspace{1cm}

{\bf Haximjan Abdusattar},
{\bf Hoernisa Iminniyaz}\footnote{Corresponding author, wrns@xju.edu.cn}

\vskip 0.15in
{\it
%$^a$
{School of Physics Science and Technology, Xinjiang University, \\
Urumqi 830046, China} \\

}

\abstract{Relic abundance of asymmetric Dark Matter
  particles in brane world cosmological scenario is investigated in this
  article. Hubble expansion rate is enhanced in brane world cosmology and
  it affects the relic abundance of asymmetric Dark Matter particles. We 
  analyze how the relic abundance of asymmetric Dark
  Matter is changed in this model. We show that in
  such kind of nonstandard cosmological scenario, indirect detection of
  asymmetric Dark Matter is possible if the cross section is small enough
  which let the anti--particle abundance kept in the same amount with the
  particle. We show the indirect detection signal constraints
  can be used to such model only when the cross section and the 5 dimensional
  Planck mass scale are in appropriate values. }
\end{center}
\end{titlepage}
\setcounter{footnote}{0}

\section{Introduction}

Recent analysis show the possibility that the Dark Matter can be
asymmetric. Most recent papers discussed the asymmetric
Dark Matter models \cite{adm-models,frandsen}. The motivation for considering
this scenario is the comparable average density of Baryon and Dark Matter
which is $\Omega_{\rm DM} = 5 \Omega_{\rm b} $ with $\Omega_{\rm b} \approx 0.046 $.
Generally neutral, long--lived or stable Weakly Interacting Massive Particles
(WIMPs) are assumed to be good Dark Matter candidates. One example is the
neutralino which is stable particle with R-parity introduced in
supersymmetry. Neutralino is Majorana
particle for which its particle and anti--particle are the same. However,
until now there is no evidence that the Dark Matter particles should be
Majorana particle. In the universe, most of known elementary particles
are not Majorana particles. The particles and anti--particles are distinct
from each other if particles are fermionic. Therefore it is legitimate to
consider the possibility that the Dark Matter particles can be asymmetric
for which particles and anti--particles are not identical.

In \cite{GSV,Iminniyaz:2011yp}, the authors investigated the
relic abundance of asymmetric Dark Matter particles in the standard
cosmological scenario in which particles were in thermal equilibrium
in the early universe and decoupled when they were non--relativistic.
Usually it is assumed the Dark Matter asymmetry is created well before Dark
Matter annihilation reactions freeze--out and in the beginning there are more
particles than the anti--particles. In the standard cosmological
scenario, the particles and anti--particles are annihilated away when the 
asymmetry is large enough. Therefore the relic density for anti--particles 
is depressed in the present universe. There is only particles. Thus the 
asymmetric Dark Matter is only assumed to be detected by direct detection.

In nonstandard cosmological scenarios like quintessence, scalar--tensor
models, brane world cosmological scenarios, the Hubble expansion rate is
increased \cite{Salati:2002md,Catena,Abou El Dahab:2006wb,Okada:2004nc}. The
modified Hubble expansion rate affects the relic density of
asymmetric Dark Matter. The asymmetric Dark Matter relic density is already
investigated in such models including
quintessence model, scalar--tensor model
 \cite{Iminniyaz:2013cla, Gelmini:2013awa, Wang:2015gua}.
In quintessence model, scalar--tensor models, the relic densities of both
particles and anti--particles are increased simultaneously due to the
increased Hubble expansion rate. The relic density of anti--particle is not
depressed for appropriate annihilation cross section for nonstandard
cosmological scenarios. This leaves the possibility
that the asymmetric Dark Matter can be detected by indirect detection.

Recently Meehan et al \cite{Meehan:2014zsa} probed the relic density of
asymmetric Dark Matter in brane world cosmology. They found the enhanced
Hubble expansion rate in brane world cosmology leads to the earlier decay of 
particle and antiparticle. Therefore there are enhanced relic densities for
both particles
and antiparticles. 
%They used WMAP data and indirect detection constraints from
%Fermi--LAT in this scenario \cite{Lahav:2014vza,Ackermann:2011wa}. 
In our work we give more detailed analysis of the relic density of asymmetric 
Dark Matter in brane world cosmology in different way. We closely follow
the analytic solution of asymmetric Dark Matter in standard cosmological
scenario and derived the analytic solution of the relic density of asymmetric
Dark Matter in brane world cosmological model.

The paper is arranged as follows. In section 2, the asymmetric Dark Matter
relic density is calculated numerically and analytically in brane world 
cosmology. In section 3, we find the constraints on the parameter space
of brane world cosmology from the observational data. The final section is
devoted to the conclusions.

\section{Relic Abundance of Asymmetric Dark Matter in Brane World Cosmology}

In this section, we calculate the relic abundance of asymmetric Dark Matter
in brane world cosmological scenario. Before starting the calculation let us
briefly review the brane world cosmological scenario. In brane world
cosmology, the Hubble expansion rate is derived as
\begin{equation}
      H^2 = \frac{8\pi G}{3} \rho \left(1 + \frac{\rho}{2\sigma}\right)
      - \frac{k}{R^2} + \frac{C^{\prime}}{R^4},
\end{equation}
where $\rho$ is the energy density of ordinary matter
on the brane with $\sigma$ being the brane tension. $R$ is the scale factor.
$G$ is 4 dimensional Newton coupling constant and $k$ is the curvature of the
3 dimensional space. Finally $C^{\prime}$ is a constant of integration which
is known as dark radiation.
We set $k=C^{\prime}=0$ in this analysis and insert $\rho = \pi^2/30 g_* T^4$ and
$\sigma = 6 M^6_5/M_{\rm Pl}^2$ into above equation where $M_5$ is the 5
dimensional Plank mass and $M_{\rm Pl}$ is the reduced Planck mass
$M_{\rm Pl} = 1/{\sqrt{8 \pi G}} = 2.4\times 10^{18}$ GeV. Then one obtains
\begin{equation}
      H = \sqrt{1 + \frac{k_b}{x^4}} \, H_{\rm std}.
\end{equation}
where $x = m_{\chi}/T$ with $m_{\chi}$ being the mass of Dark Matter particles
and $ k_b = \pi^2 g_* m_{\chi}^4 M^2_{\rm Pl}/(360 M^6_5) $. Here $g_*$ is the
effective number of the relativistic degrees of freedom which is give by
\begin{eqnarray}
  g_{*} = \sum_{i = {\rm bosons}} g_i \left( \frac{T_i}{T} \right)^4
  + \frac{7}{8} \sum_{i = {\rm fermions}} g_i
  \left( \frac{T_i}{T} \right)^4\, .
\end{eqnarray}
Here $g_i$ is the number of the internal degrees of freedom of the particles.
In order not to conflict with Big Bang Nucleosynthesis (BBN) when temperature
 $T = 1$ MeV, the second term in the square root $k_b / x^4 \to 1$. This
impose that $M_5 > 1.2 \times 10^4$ GeV. $H_{\rm std}$ is the Hubble expansion
rate in the standard cosmological scenario.

\begin{equation}
      H_{\rm std} = \sqrt{\frac{g_*}{90}}\frac{\pi m^2 }{M_{\rm Pl} \, x^2}.
\end{equation}
Following we write down the Boltzmann equation with the modified
expansion rate in brane world cosmology which is the main
equation to find the relic density for Dark Matter. In our analysis, Dark
Matter particle is denoted as  $\chi$ which is {\em not} self--conjugate,
i.e. the anti--particle $\bar\chi \neq \chi$. The Boltzmann equations for
particle and anti--particle are
\begin{eqnarray} \label{eq:boltzmann_n}
\frac{{\rm d}n_{\chi}}{{\rm d}t} + 3 H n_{\chi} &=&  - \langle \sigma v\rangle
  (n_{\chi} n_{\bar\chi} - n_{\chi,{\rm eq}} n_{\bar\chi,{\rm eq}})\,;
  \nonumber \\
\frac{{\rm d}n_{\bar\chi}}{{\rm d}t} + 3 H n_{\bar\chi} &=&
   - \langle \sigma v\rangle (n_{\chi} n_{\bar\chi} - n_{\chi,{\rm
       eq}} n_{\bar\chi,{\rm eq}})\,.
\end{eqnarray}
Here it is assumed that only $\chi \bar \chi$ pairs can annihilate into
Standard Model (SM) particles, while $\chi\chi$ and $\bar \chi \bar\chi$
pairs can not. $n_{\chi}$, $n_{\bar\chi}$ are the number densities of particle and
anti--particle. $\langle \sigma v \rangle$ is the thermally averaged
annihilation cross section multiplied with the relative velocity of the two
annihilating $\chi$, $\bar{\chi}$ particles.
$n_{\chi,{\rm eq}}$, $n_{\bar{\chi},{\rm eq}}$ are the equilibrium number
densities of $\chi$ and $\bar{\chi}$, here it is assumed the Dark Matter
particles were non--relativistic at decoupling. The equilibrium number
densities $n_{\chi,{\rm eq}}$ and $n_{\bar\chi,{\rm eq}}$ are given by
\begin{eqnarray} \label{n_eq}
  n_{\chi,{\rm eq}} &=& g_\chi ~{\left( \frac{m_\chi T}{2 \pi} \right)}^{3/2}
  {\rm e}^{(-m_\chi + \mu_\chi)/T}\,, \nonumber \\
  n_{\bar\chi,{\rm eq}} &=&  g_\chi ~{\left( \frac{m_\chi T}{2 \pi}
    \right)}^{3/2} {\rm e}^{(-m_\chi - \mu_{\bar\chi})/T}\,,
\end{eqnarray}
where $\mu_\chi$, $\mu_{\bar\chi}$ are the chemical potential of the
particle and anti--particle, $\mu_{\bar\chi} = -\mu_\chi$ in equilibrium.

The number densities for particle and anti--particle are obtained by solving
the Boltzmann equations (\ref{eq:boltzmann_n}). To solve the Boltzmann
equations (\ref{eq:boltzmann_n}) in brane world
cosmology, we follow the standard picture of Dark Matter evolution. It is
assumed the particles $\chi$ and $\bar{\chi}$ were in thermal equilibrium at
high temperature. The equilibrium number densities decrease
exponentially when temperature drops below the mass of the particle, for
 $m_\chi > |\mu_\chi|$. Finally the  falls of the interaction rates
$\Gamma = n_{\chi} \langle \sigma v \rangle$ for particle and
$\bar{\Gamma} = n_{\bar\chi} \langle \sigma v \rangle $ for anti--particle
below the Hubble expansion rate $H$ lead to the decoupling of the particles
and anti--particles from the equilibrium. The number densities of the particles
and anti--particles are almost constant from the decoupling point. The
temperature at that point is called freeze--out temperature.

First, we rewrite the Boltzmann equations (\ref{eq:boltzmann_n}) in terms of
dimensionless new quantities $Y_\chi =n_\chi/s$,
$Y_{\bar\chi} = n_{\bar\chi}/s$ and $x = m_\chi/T$.
The entropy density is given by $ s= (2 \pi^2/45) g_{*s} T^3 $,
where
\begin{eqnarray}
  g_{*s} = \sum_{i = {\rm bosons}} g_i \left( \frac{T_i}{T} \right)^3
  + \frac{7}{8} \sum_{i = {\rm fermions}} g_i
  \left( \frac{T_i}{T} \right)^3\, .
\end{eqnarray}

During the radiation dominated period, we assume that the universe expands
adiabatically, then the Boltzmann equations (\ref{eq:boltzmann_n}) become
\begin{equation} \label{eq:boltzmann_Y}
\frac{d Y_{\chi}}{dx} =
      - \frac{\lambda \langle \sigma v \rangle}{\sqrt{x^4 +k_b}}~
     (Y_{\chi}~ Y_{\bar\chi} - Y_{\chi, {\rm eq}}~Y_{\bar\chi, {\rm eq}}   )\,;
\end{equation}
\begin{equation} \label{eq:boltzmann_Ybar}
\frac{d Y_{\bar{\chi}}}{dx}
= - \frac{\lambda \langle \sigma v \rangle}{\sqrt{x^4 +k_b}}~
 (Y_{\chi}~Y_{\bar\chi} - Y_{\chi, {\rm eq}}~Y_{\bar\chi, {\rm eq}} )\,,
\end{equation}
where
\begin{equation} \label{lambda}
\lambda = 1.32\,m_{\chi} M_{\rm Pl}\, \sqrt{g_*}\,.
\end{equation}
Here it is assumed $g_* \simeq g_{*s}$ and ${\rm d}{g_*}/{\rm d}x \simeq 0$.
From these two equations (\ref{eq:boltzmann_Y}) and (\ref{eq:boltzmann_Ybar}),
we obtain
\begin{equation} \label{eq:YYbar}
\frac{d Y_{\chi}}{dx} - \frac{d Y_{\bar{\chi}}}{dx} = 0\,.
\end{equation}
This implies
\begin{equation}  \label{eq:c}
 Y_{\chi} - Y_{\bar\chi} = C\,,
\end{equation}
with C being the constant. It means the difference of the co--moving densities
of the particles and anti--particles is conserved. Using
Eq.(\ref{eq:c}), the Boltzmann equations (\ref{eq:boltzmann_Y}) and
(\ref{eq:boltzmann_Ybar}) become
\begin{equation} \label{eq:Yc}
\frac{d Y_{\chi}}{dx} =
                - \frac{\lambda \langle \sigma v \rangle}{\sqrt{x^4 + k_b}}~
                 (Y_{\chi}^2 - C Y_{\chi} - P     )\, ;
\end{equation}
\begin{equation} \label{eq:Ycbar}
\frac{d Y_{\bar{\chi}}}{dx} =
                  - \frac{\lambda \langle \sigma v \rangle}{\sqrt{x^4 +k_b}}~
 (Y_{\bar\chi}^2 + C Y_{\bar\chi}  - P)\,,
\end{equation}
where
$P = Y_{\chi, {\rm eq}} Y_{\bar\chi,{\rm eq}} = (0.145g_{\chi}/g_*)^2\, x^3 e^{-2x}$.
Mostly we use the WIMP annihilation cross section which is expanded in the
relative velocity $v$ between the annihilating WIMPs. Its thermal average is
\begin{equation} \label{eq:cross}
   \langle \sigma v \rangle = a + 6\,b x^{-1} + {\cal O}(x^{-2})\, .
\end{equation}
Here $a$ is the dominant contribution to $\sigma v$ when particle and
anti--particle are in an $s$--wave for the limit $v\to 0$. $b$ is the
$p$--wave contribution to $\sigma v$ when the $s$--wave annihilation is
suppressed.

\begin{figure}[h!]
  \begin{center}
    \hspace*{-0.5cm} \includegraphics*[width=8cm]{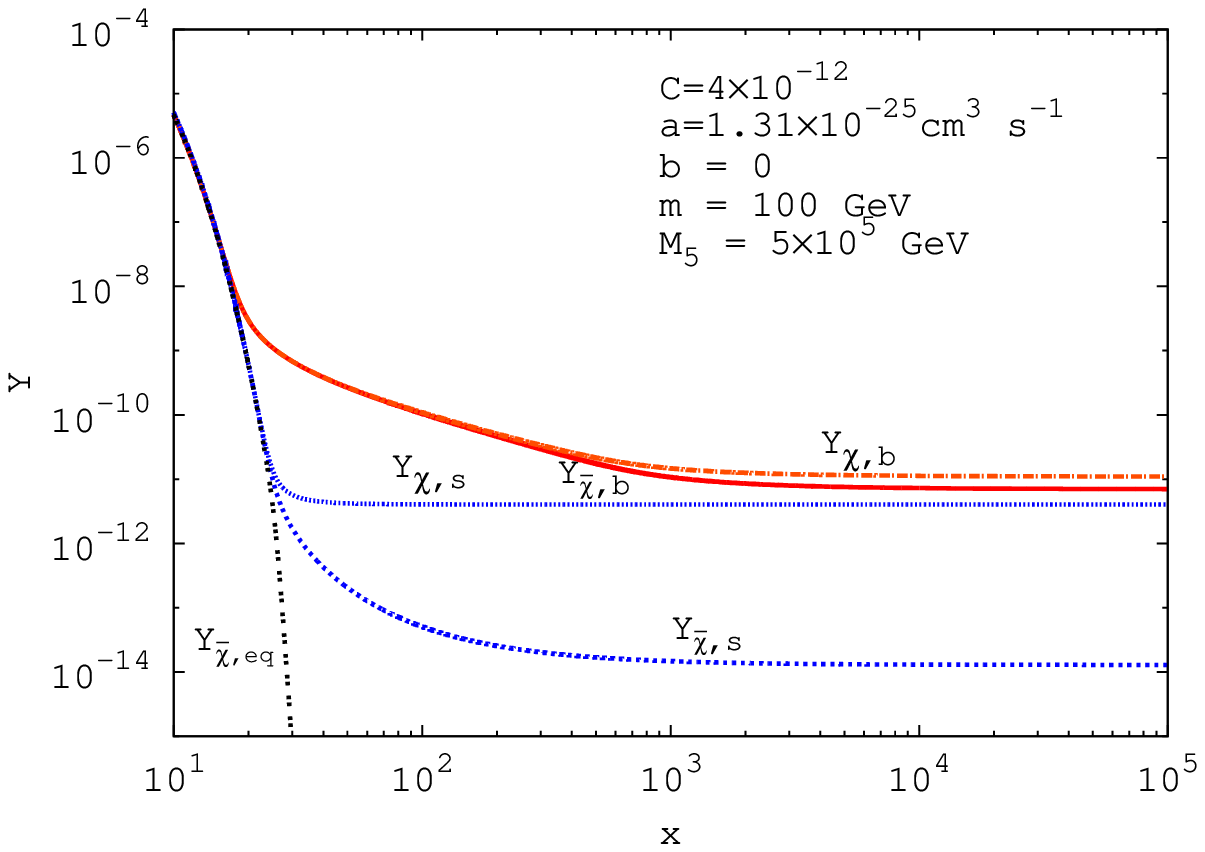}
    \put(-115,-12){(a)}
    \hspace*{-0.5cm} \includegraphics*[width=8cm]{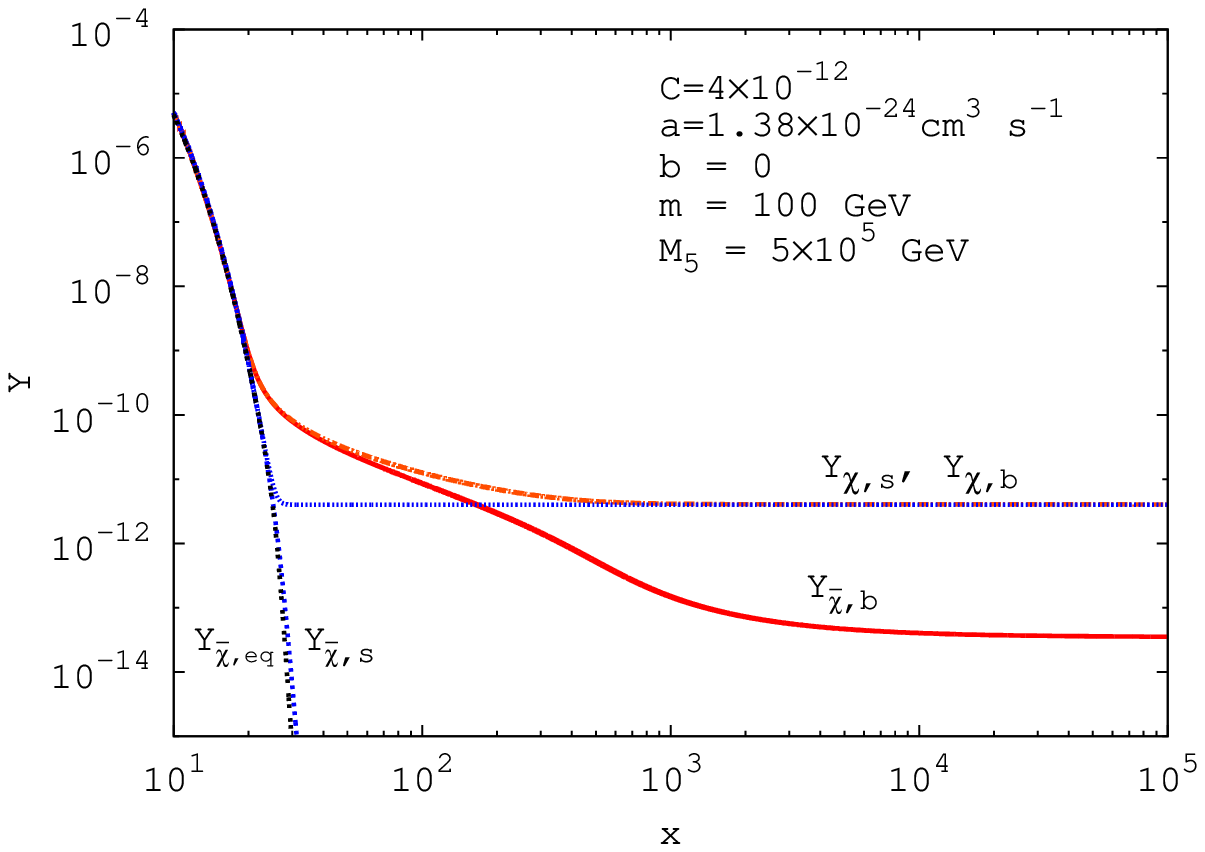}
    \put(-115,-12){(b)}
     \vspace{0.5cm}
    \hspace*{-0.5cm} \includegraphics*[width=8cm]{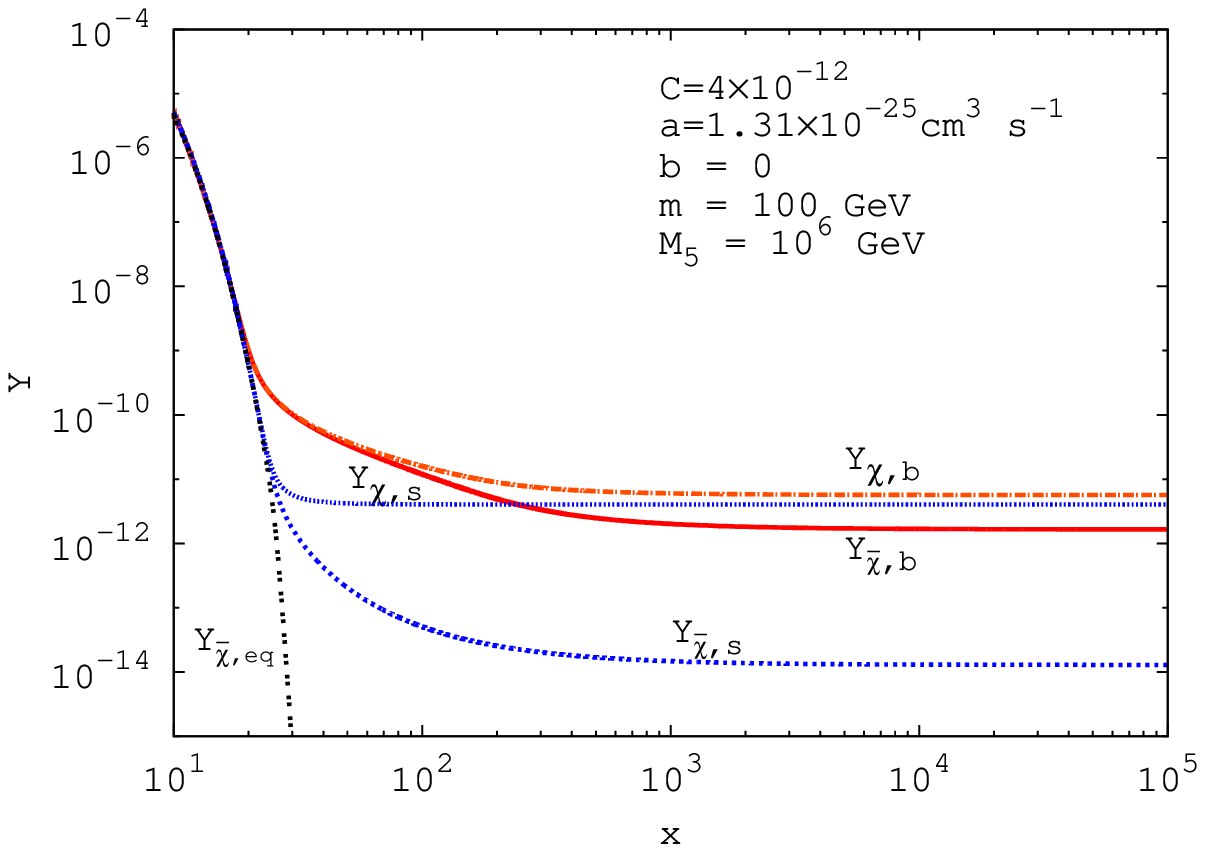}
    \put(-115,-12){(c)}
    \hspace*{-0.5cm} \includegraphics*[width=8cm]{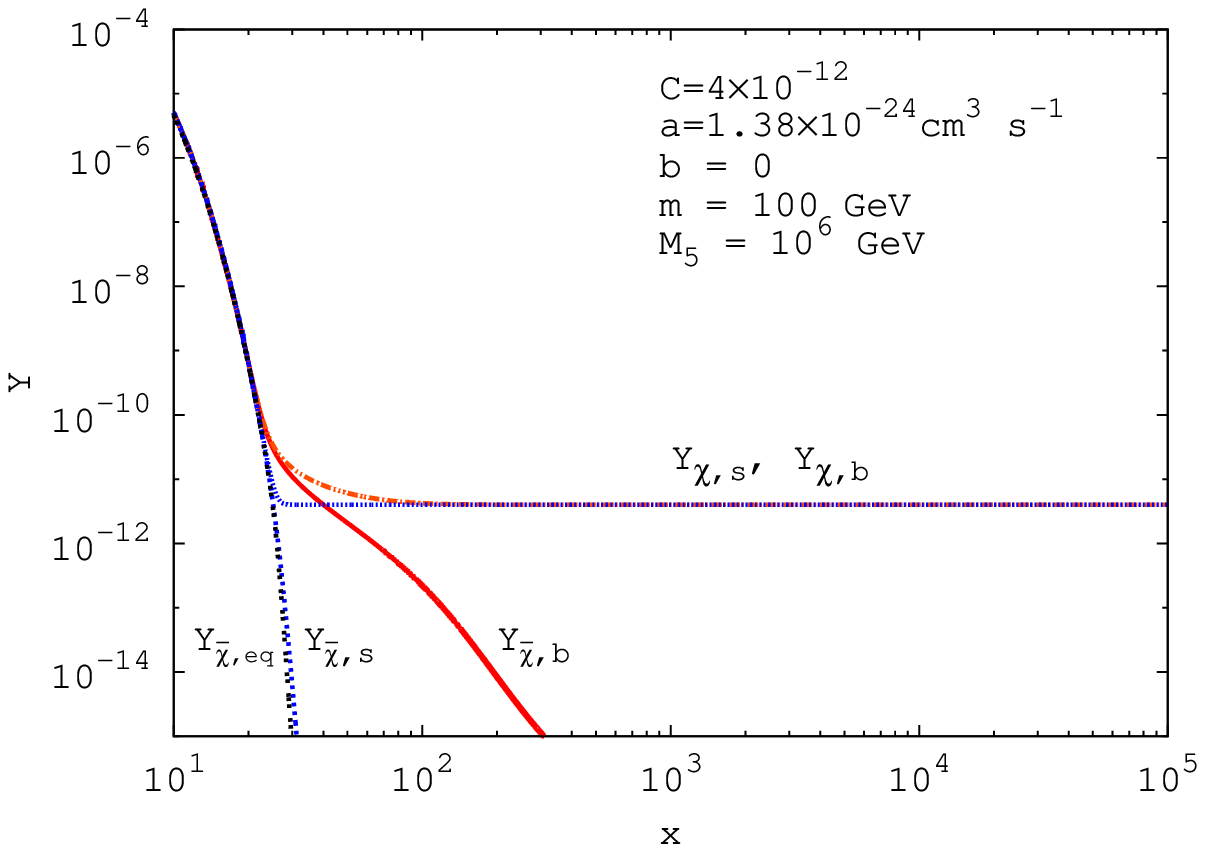}
    \put(-115,-12){(d)}
    \vspace{0.3cm}
    \hspace*{-0.5cm} \includegraphics*[width=8cm]{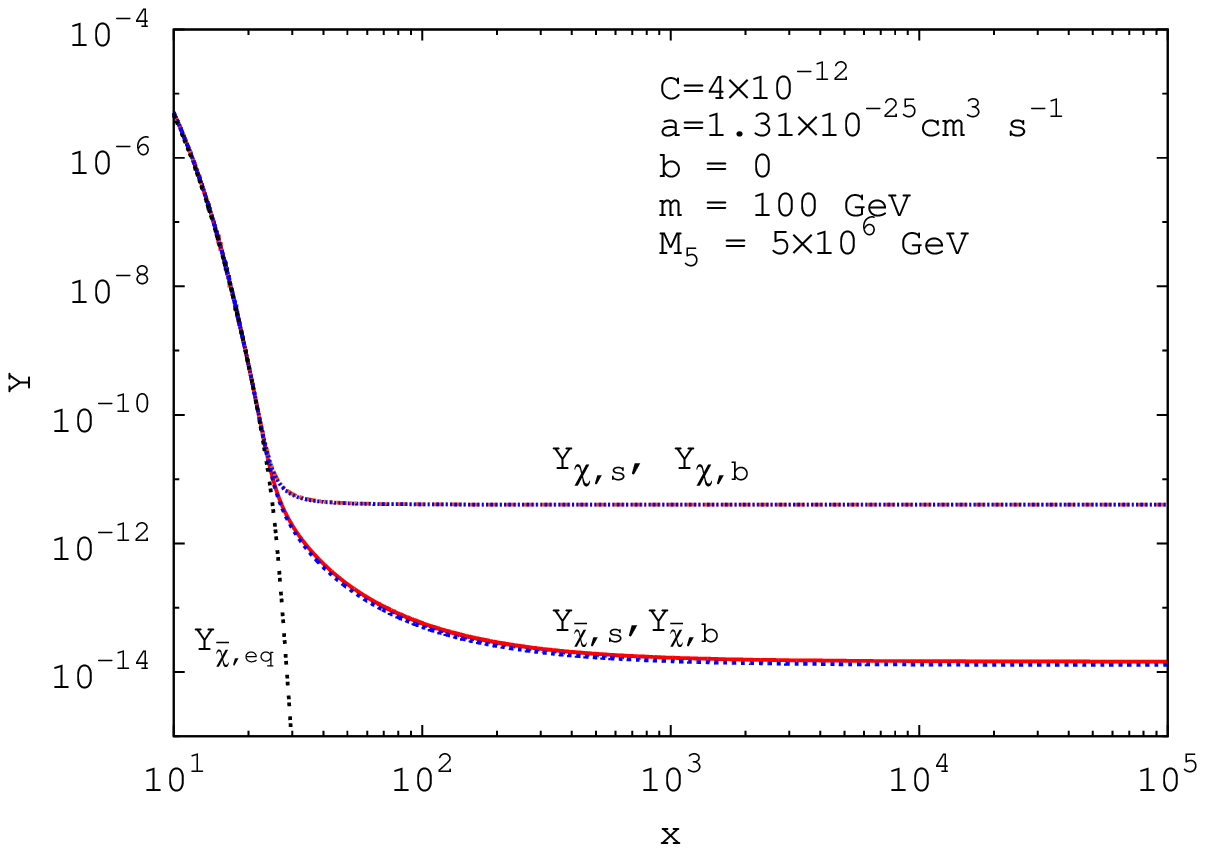}
    \put(-115,-12){(e)}
    \hspace*{-0.5cm} \includegraphics*[width=8cm]{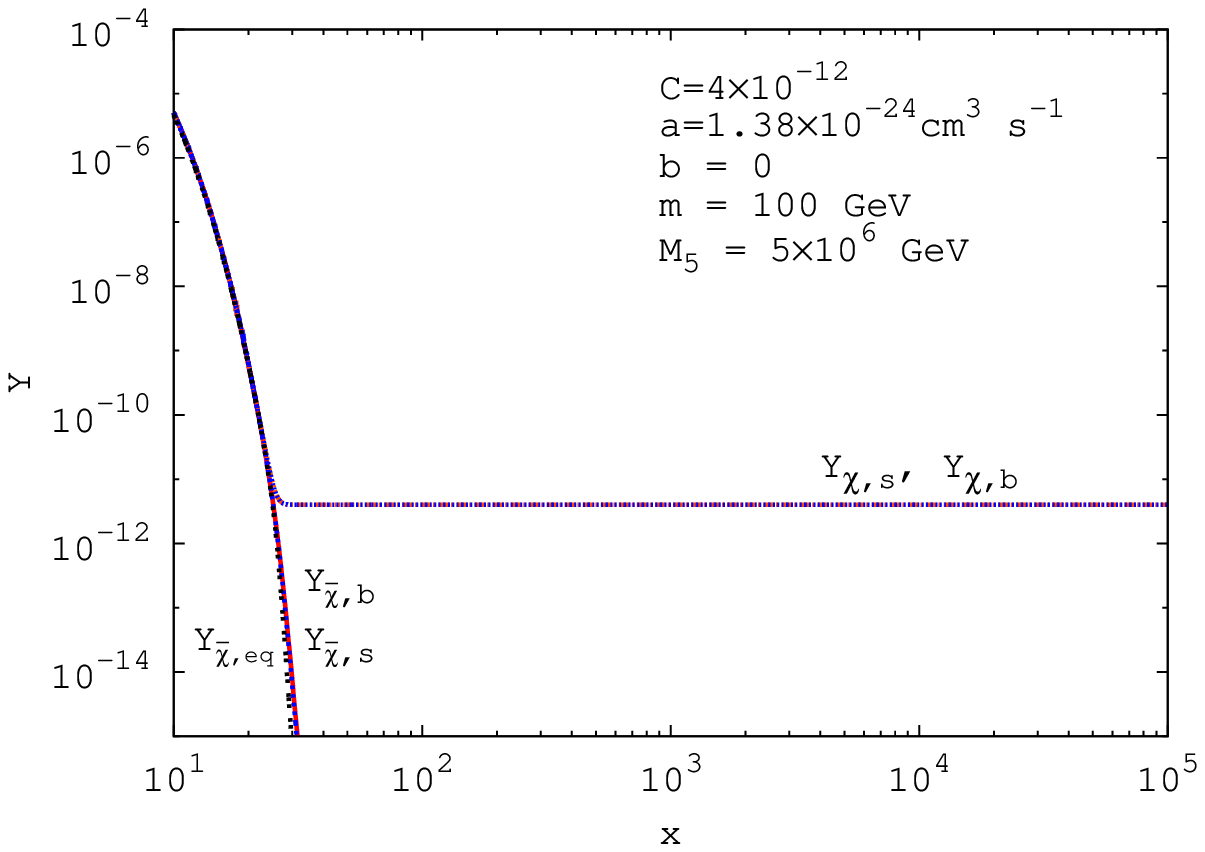}
    \put(-115,-12){(f)}
    \caption{\label{fig:a} \footnotesize The relic abundances $Y_{\chi}$ and
     $Y_{\bar\chi}$ for particle and anti--particle  as a
     function of the inverse--scaled temperature $x$ for
     $ C = 4 \times 10^{-12} $, $m_\chi = 100$ GeV, $g_{\chi} = 2$,
     $g_* = 90$, $ a = 1.31 \times 10^{-25} $cm$^3$ s$^{-1}$, $b = 0$ for
     panels (a), (c), (e) and $ a = 1.38 \times 10^{-24} $ cm$^3$ s$^{-1}$,
     $b = 0$ for panels (b), (d), (f); $M_5 = 5 \times 10^5$ GeV in (a), (b);
     $M_5 = 10^6$ GeV in (c), (d) and
     $M_5 = 5 \times 10^6$ GeV in (e), (f). }
  \end{center}
\end{figure}
Fig.\ref{fig:a} shows the relic abundance $Y_{\chi, \bar\chi}$ evolution as a
function of the inverse--scaled temperature $x$ for particle and
anti--particle for different cross sections and $M_5$ in brane
cosmology scenario and standard scenario. This figure is plotted using
the numerical solutions of
Eqs.(\ref{eq:Yc}), (\ref{eq:Ycbar}). Here we take $ C = 4 \times 10^{-12} $,
 $m_\chi = 100$ GeV, $g_{\chi} = 2$, $g_* = 90$, $ a = 1.31 \times 10^{-25} $
cm$^3$ s$^{-1}$, $b = 0$ for panels (a), (c), (e) and $ a = 1.38 \times 10^{-24} $
cm$^3$ s$^{-1}$, $b = 0$ for panels (b), (d), (f); $M_5 = 5 \times 10^5$ GeV
in (a), (b); $M_5 =  10^6$ GeV in (c), (d) and
$M_5 =  5 \times 10^6$ GeV in (e), (f).
The dot--dashed (red) line is the particle abundance $Y_{\chi,b}$ and the
thick (red) line is the anti--particle abundance $Y_{\bar\chi,b}$ in brane
world cosmology; the dashed
(blue) line is particle abundance $Y_{\chi,s}$ and dotted (blue) line is the
anti--particle abundance $Y_{\bar\chi,s}$ in the standard scenario. The double
dotted (black) line is the equilibrium value $Y_{\bar\chi,{\rm eq}}$ of the
anti--particle abundance. Hubble expansion rate in brane world cosmology is
larger than the Hubble expansion rate in the standard case. Particles and
anti--particles freeze--out earlier than the standard cosmology. After decay
of thermal equilibrium, there are more particles and anti--particles left
in brane world cosmology than the standard case for appropriate cross
section. Of course the extent of the
increase depends on the
cross sections and 5 dimensional scale $M_5$. The relic abundances $Y_{\chi,b}$
for particle $\chi$ and $Y_{\bar\chi,b}$ for anti--particle $\bar\chi$ are larger
than the standard case for panels (a) and (c) when the
cross section is $ a = 1.31 \times 10^{-25} $cm$^3$ s$^{-1}$ for different
$M_5$. We found for the same cross section the increase is larger for smaller
$M_5$. When $M_5 = 5 \times 10^6$ GeV in panel (e), the relic abundances for
particle and anti--particle in brane world cosmology are almost same with the
standard result. There is no enhancement for the relic density for particles
and anti--particles. On the other hand, in brane world cosmology, for the
larger cross section as $ a = 1.38 \times 10^{-24} $cm$^3$ s$^{-1}$ the
particle abundance is almost
kept in the same amount with the particle abundance in standard cosmology for
different scaled $M_5$. This is shown in panels (b), (d) and (f). The
anti--particle abundance is depressed for such kind of large cross section.
It is more obvious for larger $M_5$ as in panel (f).

Using the same method as \cite{Iminniyaz:2011yp}, we obtain the analytic
solution of the relic density for asymmetric Dark Matter in brane world
cosmology. We solve Eq.(\ref{eq:Ycbar}) for $\bar{\chi}$ density
first and then using the relation $Y_{\chi} - Y_{\bar\chi} = C$ to find
$Y_{\chi}$ easily. The Boltzmann equation (\ref{eq:Ycbar}) is rewritten as:
\begin{equation} \label{eq:delta}
\frac{d \Delta_{\bar\chi}}{dx} = - \frac{d Y_{\bar\chi,{\rm eq}}}{dx} -
\frac{\lambda \langle \sigma v \rangle}{\sqrt{x^4 + k_b}}~
\left[\Delta_{\bar\chi}(\Delta_{\bar\chi} + 2 Y_{\bar\chi,{\rm eq}})
      + C \Delta_{\bar\chi}   \right]\, ,
\end{equation}
where $\Delta_{\bar\chi} = Y_{\bar\chi} - Y_{\bar\chi,{\rm eq}}$.
We consider two extreme cases. At high temperature, $Y_{\bar\chi}$ closely
tracks its equilibrium value $Y_{\bar\chi,{\rm eq}}$ very well.
$\Delta_{\bar\chi}$ is small, then we can neglect $d \Delta_{\bar\chi}/dx$ and
$\Delta_{\bar\chi}^2$. The Boltzmann equation (\ref{eq:delta}) then becomes
\begin{equation}\label{eq:delta_simp}
     \frac{d Y_{\bar\chi,{\rm eq}}}{dx}   =  -
      \frac{\lambda \langle \sigma v \rangle}{\sqrt{x^4 + k_b}}~
      \left(2 \Delta_{\bar\chi} Y_{\bar\chi,{\rm eq}} +
       C \Delta_{\bar\chi} \right)\,.
\end{equation}
Repeating the same way as in \cite{Iminniyaz:2011yp}, one obtains the solution
for $\Delta_{\bar\chi}$
\begin{equation} \label{bardelta_solu}
      \Delta_{\bar\chi} \simeq \frac{2 P \sqrt{x^4 + k_b}}
      {\lambda \langle \sigma v \rangle\,(C^2 + 4 P)}\,.
 \end{equation}
This solution is used to determine the freeze--out temperature $\bar{x}_F$
for $\bar{\chi}$.

In the second case, the production term $\propto Y_{\bar\chi,{\rm eq}}$ can be
ignored in the Boltzmann equation (\ref{eq:delta}) for sufficiently low
temperature, i.e. for $x > \bar x_F$, therefore
\begin{equation} \label{eq:delta_late}
\frac{d \Delta_{\bar\chi}}{dx} = - \frac{\lambda \langle \sigma v
\rangle}{\sqrt{x^4 + k_b}} \left( \Delta_{\bar\chi}^2 + C\Delta_{\bar\chi} \right)\,.
\end{equation}
After integration of Eq.(\ref{eq:delta_late}) from
$\bar{x}_F$ to $\infty$, we obtain the final WIMP abundance. We have to be
careful that the integration range is divided into two parts: one part is from
$\bar{x}_F$ to $x_t$ where the brane world cosmology is applied. $x_t $ is the
transition temperature at which point the standard cosmology recovers. It is
given by
$x_t = m_{\chi}/(0.51 \times 10^{-9} M_5^{3/2})$ \cite{Abou El Dahab:2006wb}.
Second part is from $x_t$ to $\infty$ in which the standard cosmology is
used. Again assuming $\Delta_{\chi}(\bar{x}_F) \gg \Delta_{\chi}(\infty) $, we
have
\begin{equation} \label{eq:barY_cross}
Y_{\bar\chi}(x \rightarrow \infty) =  \frac{C}
 { \exp \left[ 1.32\, C \, m_{\chi} M_{\rm Pl}\,
 \sqrt{g_*} \,  I(\bar{x}_F, x_t)    \right] -1}\,,
\end{equation}
where
\begin{eqnarray}
  I(\bar{x}_F, x_t)   & =&
              \int^{x_t}_{\bar{x}_F} \frac{ \langle \sigma v \rangle }
              {\sqrt{x^4 + k_b}}~ dx + \int^{\infty}_{x_t}
              \frac{ \langle \sigma v \rangle }
              {x^2}~ dx \\ &=& a \left( \frac{1}{\bar{x}_F}\,
         {_2}F_1 \left[ \frac{1}{4}, \frac{1}{2}, \frac{5}{4},
          \frac{-k_b}{\bar{x}^4_F}\right] - \frac{1}{x_t}\,
         {_2}F_1 \left[ \frac{1}{4}, \frac{1}{2}, \frac{5}{4},
          \frac{-k_b}{x^4_t}\right] \right) \\
         & + & \frac{3b}{\sqrt{k_b}}\,
         \left[ {\rm sinh}^{-1} \left( \frac{\sqrt{k_b}}{\bar{x}^2_F}  ) -
         {\rm sinh}^{-1}(\frac{\sqrt{k_b}}{x^2_t} \right)   \right]
         + \frac{a}{x_t} + \frac{3b}{x_t^2}\, .
\end{eqnarray}
Using equation (\ref{eq:c}), we obtain the relic abundance
for $\chi$ particle,
\begin{equation}\label{eq:Y_cross}
      Y_{\chi}(x \rightarrow \infty) = \frac{C}
 {1 - \exp \left[- 1.32\, C \, m_{\chi} M_{\rm Pl}\,
 \sqrt{g_*} \,  I(x_F, x_t)    \right] }\,.
\end{equation}
Eqs.(\ref{eq:barY_cross}) and (\ref{eq:Y_cross}) are
consistent with the constraint (\ref{eq:c}) if $x_F = \bar x_F$.
Usually the final abundance is expressed as

\begin{equation}
\Omega_\chi h^2 =\frac{m_\chi s_0 Y_{\chi}(x \to \infty) h^2}{\rho_{\rm  crit}}\,,
\end{equation}
here $s_0 = 2.9 \times 10^3~{\rm cm}^{-3}$ is the present
entropy density, and $\rho_{\rm crit} = 3 M_{\rm Pl}^2 H_0^2$ is the present
critical density.
The prediction for the present relic density for Dark Matter is then
given by
\begin{eqnarray} \label{omega}
 \Omega_{\rm DM}  h^2 & = & 2.76 \times 10^8~ m_\chi \left[ Y_{\chi}~(x
  \rightarrow \infty) + Y_{\bar\chi}~(x \rightarrow \infty) \right] \\
          \nonumber  & = &
               \frac{2.76 \times 10^8~ m_\chi ~C}
 {1 - \exp \left[- 1.32\, C \, m_{\chi} M_{\rm Pl}\,
 \sqrt{g_*} \,  I(x_F, x_t)    \right] } + \frac{2.76 \times 10^8~ m_\chi ~C}
 { \exp \left[ 1.32\, C \, m_{\chi} M_{\rm Pl}\,
 \sqrt{g_*} \,  I(\bar{x}_F, x_t)    \right] -1} \,.
\end{eqnarray}
The freeze--out temperature for $\bar{\chi}$ is fixed using the standard
method. It is assumed that the deviation
$\Delta_{\bar\chi} $ is of the same order of the equilibrium value of
$Y_{\bar\chi}$:
\begin{equation} \label{xf1}
\xi Y_{\bar\chi,{\rm eq}}( \bar{x}_F) = \Delta_{\bar\chi}( \bar{x}_F)\,,
\end{equation}
where $\xi$ is a numerical constant of order unity. We find the approximate
analytic result matches with the exact numerical result very well when we
choose $\xi = \sqrt{2} -1$ \cite{standard-cos}.

%%%%%%%%%%%%%%%%%%%%%%%%%%%%%%%%%%%%%%%%%%%%%%%%%%%%%%%%%%%%%%%%%%%%%

\section{Constraints on Parameter Space}
\setcounter{footnote}{0}

Nine-year Wilkinson Microwave Anisotropy Probe (WMAP) observations
give the  Dark Matter relic density as \cite{wmap}
\begin{eqnarray} \label{wmap}
  \Omega_{\rm DM} h^2 = 0.1138 \pm 0.0045\, ,
\end{eqnarray}
where $\Omega_{\rm DM}$ is the Dark Matter (DM) density in units of
the critical density, and $h = 0.738 \pm 0.024$ is the Hubble
constant in units of 100 km s$^{-1}$ Mpc$^{-1}$.
We use this result to find
the constraints on the parameters in brane world cosmology for asymmetric Dark
Matter. We use the following range for Dark Matter relic
density,
\begin{equation} \label{range}
0.10 < \Omega_{\rm DM} h^2 < 0.12
\end{equation}
The particle $\chi$ and anti--particle $\bar{\chi}$ density contribute
together to the total Dark Matter density:
\begin{equation} \label{add}
\Omega_{\rm DM}=\Omega_\chi + \Omega_{\bar{\chi}}\,.
\end{equation}
\begin{figure}[t!]
  \begin{center}
    \hspace*{-0.5cm} \includegraphics*[width=8.7cm]{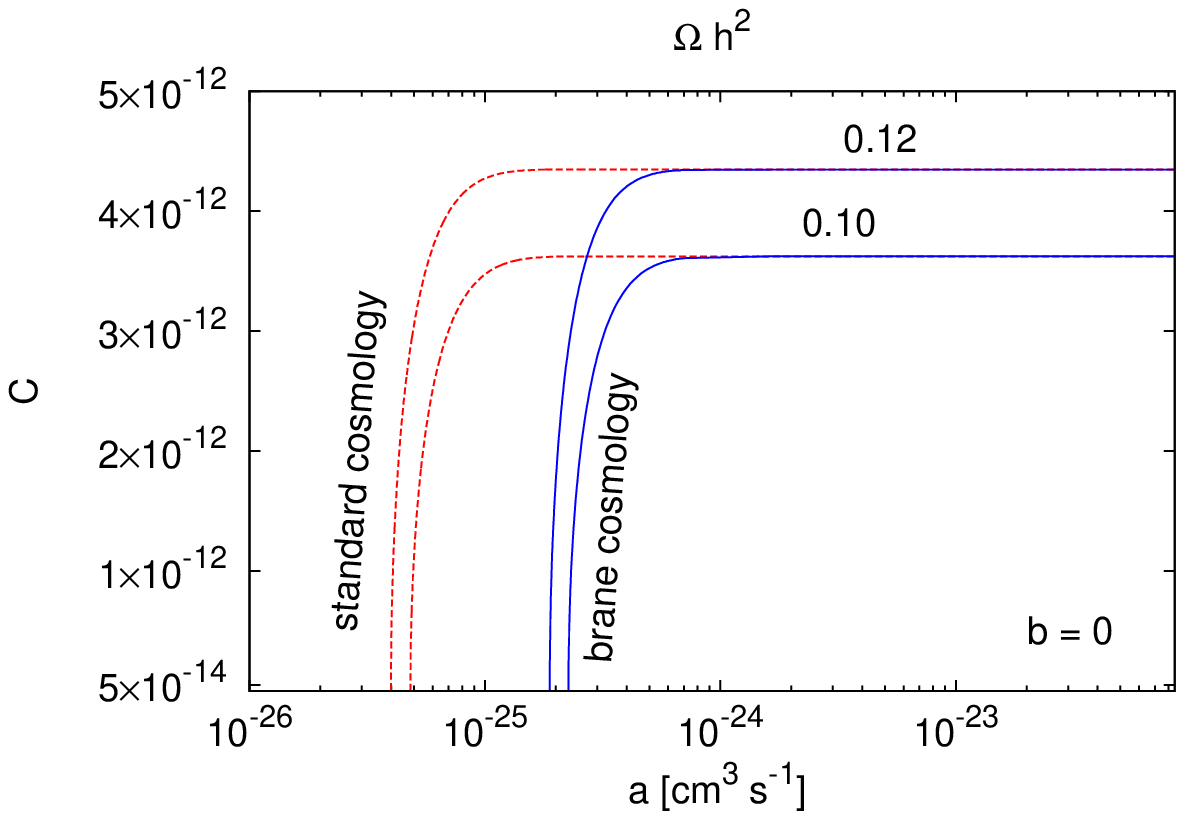}
    \put(-115,-12){(a)}
    \hspace*{-0.5cm} \includegraphics*[width=8.7cm]{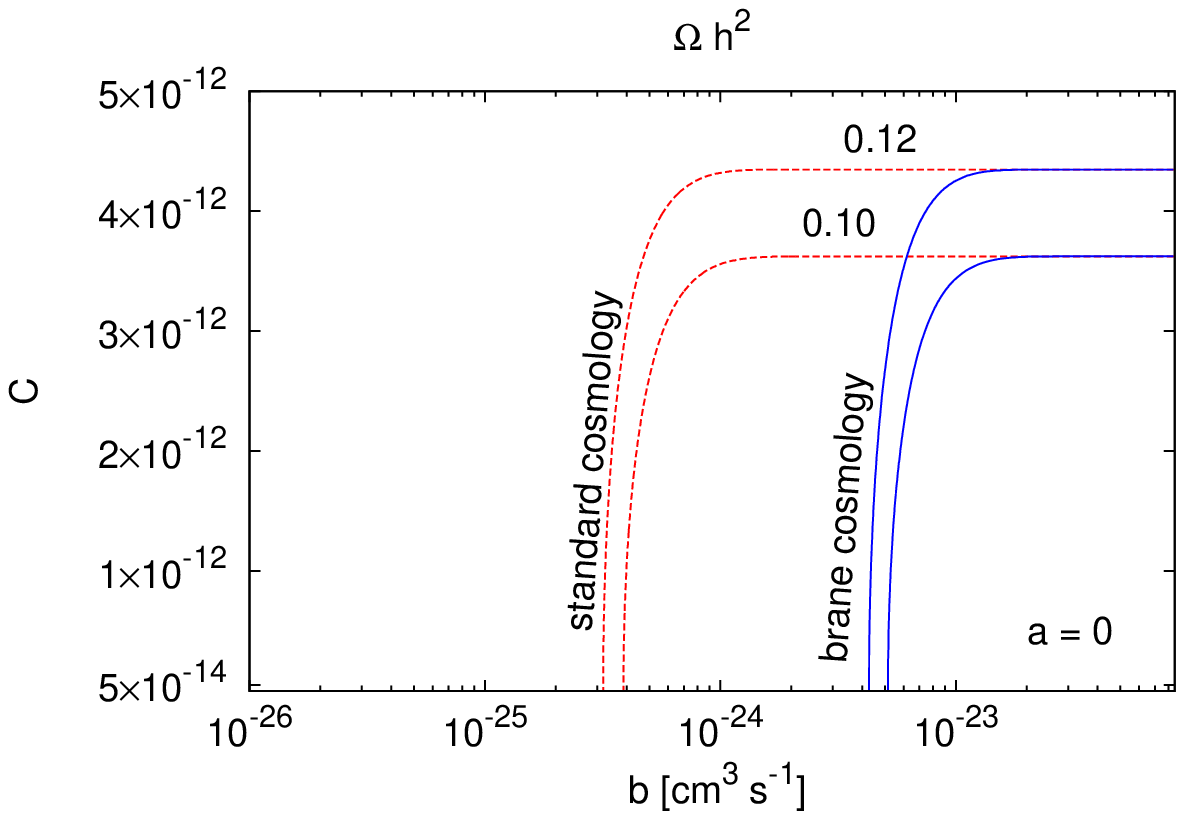}
    \put(-115,-12){(b)}
    \caption{\label{fig:d} \footnotesize
    The allowed region in the $(a,C)$ plane for $b=0$ (left), and $(b,C)$
    plane for $a=0$ (right) when the Dark Matter density $\Omega h^2$ lies
    between 0.10 and 0.12. Here we
    take $m_\chi = 100$ GeV, $g_{\chi} = 2$ and $g_* = 90$, $M_5 = 10^6$ GeV.
    The dashed (red) line is for standard cosmology and the thick (blue) line
    is for brane cosmology.}
    \end{center}
\end{figure}
Fig.\ref{fig:d} shows the relation between the cross section parameters
$a$, $b$ and asymmetry factor $C$ for the brane world cosmology and standard
cosmology when the Dark Matter relic density $\Omega_{\rm DM} h^2$ lies
between 0.10 and 0.12.
Here we take $m_\chi = 100$ GeV, $g_{\chi} = 2$ and $g_* = 90$,
$M_5 = 10^6$ GeV. The dashed (red) lines are for the standard
cosmology and the thick (blue) lines are for the brane world cosmological
scenario. When $C$ is small
as $C = 0$ to $C = 3.5 \times 10^{-5}$, the cross section is almost
independent of $C$. The reason is that for small $C$, the formulae for
$Y_{\bar\chi}$, Eq.(\ref{eq:barY_cross}) and $Y_{\chi}$, Eq.(\ref{eq:Y_cross}) 
can be recovered to the result for symmetric Dark Matter in standard 
cosmology. When the
asymmetry $C$ is large, e.g. $C = 4.0 \times 10^{-12}$, the particle and
anti--particle are completely annihilated away and the relic density is
independent of the cross section. On the other hand, for brane
world cosmology, the cross section is larger than the standard cosmology for
the observed value of Dark Matter. The reason is that in brane world
cosmological scenario the interaction rate
is larger than the standard scenario. Asymmetric Dark Matter particles
need larger annihilation cross sections in order to fall in the observation
range. In the left panel (a) of Fig.\ref{fig:d}, it is shown for s--wave
annihilation cross sections from
$a = 3.99 \times 10^{-26}$ ${\rm cm}^3 {\rm s}^{-1}$ to
$a = 8.50 \times 10^{-23}$ ${\rm cm}^3 {\rm s}^{-1}$, the observed Dark
Matter abundance is obtained  for
$C = 0$ to $4.3 \times10^{-12}$. It is for
the standard cosmology. The initial cross--section is increased to
$a = 1.93 \times 10^{-25}$ ${\rm cm}^3 {\rm s}^{-1}$ for the brane world
cosmology. The similar behavior is seen in the right panel (b) of
Fig.\ref{fig:d}.
The cross section is increased evidently for brane world cosmology for
both s--wave annihilation and p--wave annihilation cross sections to obtain
the observed Dark Matter abundance.

For asymmetric Dark Matter, since in the beginning it is assumed that there is
less anti--particle than the particle and at late time the anti--particle is
completely annihilated away with the particle. Therefore the rest is the
particle which made up the present total Dark Matter abundance in asymmetric 
Dark Matter case in standard cosmological scenario. Thus the asymmetric Dark
Matter particle is supposed to be detected by direct detection. Usually the
indirect detection is not available for the asymmetric Dark Matter.
However, for asymmetric Dark Matter in nonstandard cosmological
scenarios including quintessence, scalar--tensor model and brane world
cosmology \cite{Iminniyaz:2013cla,Gelmini:2013awa,Wang:2015gua},
the relic densities for both particle and anti--particle are increased in
due to the enhanced Hubble expansion
rate. Again for appropriate cross section, the relic density for both particle
and anti--particle are almost in the
same amount which is different from the standard
cosmological scenario where the anti--particle relic density is depressed .
Therefore the indirect detection is possible for nonstandard cosmology.
However, for large annihilation cross section as 
$a = 1.38 \times 10^{-24}$ ${\rm cm}^3$ ${\rm s}^{-1}$, the indirect detection 
sigal can not be used. In
Fig.\ref{fig:a}, it is shown when the
cross section is as large as
$a = 1.38 \times 10^{-24}$ ${\rm cm}^3$ ${\rm s}^{-1}$, the relic
abundance of anti--particle is suppressed. Therefore
there is not enough anti--particle left which can annihilate with particle to
make up the indirect detection signal. Thus the indirect detection for
asymmetric Dark Matter is available in conditionally where the annihilation
cross section is not so large. As we showed in Fig.\ref{fig:a}, it depends on
the scale of $M_5$, the mass of Dark Matter $m_{\chi}$, and most important is
the cross section. In this article, it is assumed
there is only particle and anti--particle annihilations. Therefore the
abundant Dark Matter particle can not annihilate with
itself to make the product particle which can be detected by indirect 
detection experiments.

\begin{figure}[t!]
  \begin{center}
    \hspace*{-0.5cm} \includegraphics*[width=8.7cm]{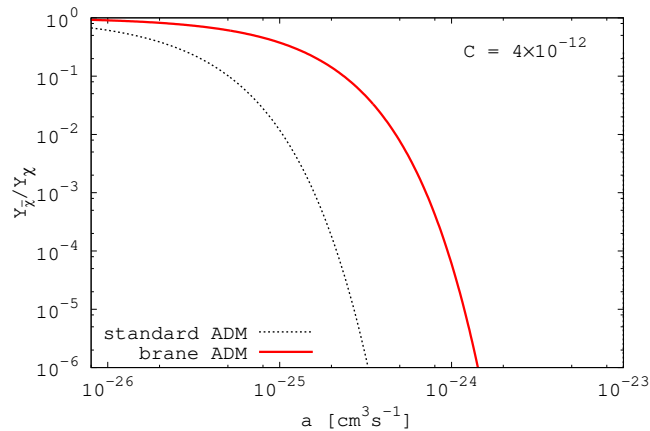}
%    \put(-115,-12){(a)}
    \caption{\label{fig:ybary} \footnotesize
    Ratio of the relic abundance of anti--particle and particle as a function 
    of annihilation cross section $a$ in brane world cosmology and in the 
    standard cosmology for the asymmetry factor $C = 4 \times 10^{-12}$. The 
    dotted (black) line is for the asymmetric Dark Matter (ADM) in standard
    cosmology and the solid (red) line is for the asymmetric Dark Matter in
    brane world cosmology. Here $b=0$, $m_\chi=100$ GeV and
   $C = 4 \times 10^{-12}$, $M_5 = 10^6$ GeV, $g_{\chi} = 2$, $g_* = 90$. }
    \end{center}
\end{figure}

The ratio of the relic abundance of the anti--particle to the particle as a 
function of the annihilation cross section $a$ is plotted in 
Fig.\ref{fig:ybary} for $C = 4 \times 10^{-12}$. Here the dotted (black) line 
for asymmetric Dark Matter in standard cosmology and the solid (red) line in 
brane world cosmology. The decrease of the ratio is slower for brane world 
cosmology comparing to the standard cosmology as the cross section increases. 
For example, for the cross section value 
$a = 2.50 \times 10^{-25}$ ${\rm cm}^3$ ${\rm s}^{-1}$, the ratio 
is decreased to $2.2\times 10^{-5}$ for standard cosmology and 
$8.6\times 10^{-2}$ for brane world 
cosmology. We noticed that when the cross section increases, in both cases the 
abundances of anti--particles are suppressed. Therefore the indirect detection 
signal is not used for too large cross section in brane world cosmoloy for 
asymmetric Dark Matter.

\section{Summary and Conclusions}

In this paper we investigated the relic abundance of asymmetric WIMP
Dark Matter in brane world cosmology. For asymmetric Dark Matter the particles
and anti--particles are different from each other. It is assumed the
asymmetry starts well before the epoch of thermal decoupling of the WIMPs. The
asymmetric Dark Matter particles and anti--particles freeze--out earlier
than the standard cosmology due to the enhanced Hubble rate in brane world
cosmology. This leads to the increase of the relic density of asymmetric Dark
Matter particles.

We found that the relic densities of both particles and
anti--particles are increased in brane world cosmology. The size of the
increase depends on the 5 dimensional Planck mass scale $M_5$ and the cross
sections. The increases are more sizable for smaller $M_5$ for the same cross
section. When $M_5$ is large, e.g. $M_5 = 5\times 10^6$ GeV, the case in the 
standard cosmology is recovered. For larger annihilation cross section as
$a = 1.38 \times 10^{-24} {\rm cm^3}$ $ {s^{-1}}$, the particle abundance in brane
world cosmology is the same with the particle abundance in standard cosmology
for different $M_5$. The antiparticle abundance is suppressed.

We used the WMAP data to give the constraints on the cross section and
asymmetry factor $C$ in brane world cosmology. We showed that the indirect
detection signal is conditionally used for asymmetric Dark Matter in brane
world cosmology.

The result which is obtained in our work is important to understand the relic
abundance of asymmetric Dark Matter in the early universe before Big Bang
Nucleosynthesis. If we know the annihilation cross section in the brane 
world cosmology, we can find the relic density of Dark Matter in this
scenario. It has important effect to investigate the early universe before BBN.

\section*{Acknowledgments}

The work is supported by the National Natural Science Foundation of China
(11365022).


\begin{thebibliography}{99}

\bibitem{adm-models}
S.~Nussinov, Phys.\ Lett.\ B {\bf 165}, 55 (1985);
K.~Griest and D.~Seckel.
%Cosmic Asymmetry, Neutrinos and the Sun.
Nucl. Phys. B {\bf 283}, 681 (1987);
R.~S.~Chivukula and T.~P.~Walker, Nucl.\ Phys.\ B {\bf 329}, 445 (1990);
D.~B.~Kaplan, Phys.\ Rev.\ Lett.\ {\bf 68}, 742 (1992);
D.~Hooper, J.~March-Russell and S.~M.~West,
  %``Asymmetric sneutrino Dark Matter and the Omega(b)/Omega(DM) puzzle,''
 Phys.\ Lett.\  B {\bf 605}, 228 (2005) [arXiv:hep-ph/0410114];
%Decaying Dark Matter can explain the electron/positron excesses
JCAP {\bf 0901} (2009) 043 [arXiv:0811.4153v1 [hep-ph]];
H.~An, S.~L.~Chen, R.~N.~Mohapatra and Y.~Zhang,
  %``Leptogenesis as a Common Origin for Matter and Dark Matter,''
  JHEP {\bf 1003}, 124 (2010) [arXiv:0911.4463 [hep-ph]];
T.~Cohen and K.~M.~Zurek,
  %``Leptophilic Dark Matter from the Lepton Asymmetry,''
  Phys.\ Rev.\ Lett.\  {\bf 104}, 101301 (2010) [arXiv:0909.2035 [hep-ph]].
D.~E.~Kaplan, M.~A.~Luty and K.~M.~Zurek,
  %``Asymmetric Dark Matter,''
  Phys.\ Rev.\  D {\bf 79}, 115016 (2009) [arXiv:0901.4117 [hep-ph]];
T.~Cohen, D.~J.~Phalen, A.~Pierce and K.~M.~Zurek,
  %``Asymmetric Dark Matter from a GeV Hidden Sector,''
Phys.\ Rev.\  D {\bf 82}, 056001 (2010) [arXiv:1005.1655 [hep-ph]];
J.~Shelton and K.~M.~Zurek,
  %``Darkogenesis: A baryon asymmetry from the Dark Matter sector,''
Phys.\ Rev.\  D {\bf 82}, 123512 (2010) [arXiv:1008.1997 [hep-ph]];


\bibitem{frandsen}
A.~Belyaev, M.~T.~Frandsen, F.~Sannino and S.~Sarkar, Phys. Rev. D
{\bf 83}, 015007 (2011) [arXiv:1007.4839].

\bibitem{GSV}
M. L. Graesser, I. M. Shoemaker and L. Vecchi, [arxiv:1103.2771 [hep-ph]].


%\cite{Iminniyaz:2011yp}
\bibitem{Iminniyaz:2011yp}
  H.~Iminniyaz, M.~Drees and X.~Chen,
  %``Relic Abundance of Asymmetric Dark Matter,''
  JCAP {\bf 1107}, 003 (2011)
  [arXiv:1104.5548 [hep-ph]].
  %%CITATION = ARXIV:1104.5548;%%



\bibitem{Salati:2002md}
  P.~Salati,
  %``Quintessence and the relic density of neutralinos,''
  Phys.\ Lett.\ B {\bf 571}, 121 (2003)
  [astro-ph/0207396].
  %%CITATION = ASTRO-PH/0207396;%%

\bibitem{Catena}
R.~Catena, N.~Fornengo, A.~Masiero, M.~Pietroni and F.~Rosati,
Phys. Rev.  D {\bf 70}, 063519 (2004) [arXiv:astro-ph/0403614].

\bibitem{Abou El Dahab:2006wb}
  E.~Abou El Dahab and S.~Khalil,
  %``Cold dark matter in brane cosmology scenario,''
  JHEP {\bf 0609} (2006) 042
  [hep-ph/0607180].
  %%CITATION = HEP-PH/0607180;%%

%\cite{Okada:2004nc}
\bibitem{Okada:2004nc}
  N.~Okada and O.~Seto,
  %``Relic density of dark matter in brane world cosmology,''
  Phys.\ Rev.\ D {\bf 70} (2004) 083531
  [hep-ph/0407092].
  %%CITATION = HEP-PH/0407092;%%
  %44 citations counted in INSPIRE as of 24 Apr 2015


%\cite{Iminniyaz:2013cla}
\bibitem{Iminniyaz:2013cla}
  H.~Iminniyaz and X.~Chen,
  %``Relic Abundance of Asymmetric Dark Matter in Quintessence,''
  Astropart.\ Phys.\  {\bf 54}, 125 (2014)
  [arXiv:1308.0353 [hep-ph]].
  %%CITATION = ARXIV:1308.0353;%%
  %3 citations counted in INSPIRE as of 25 Apr 2014

%\cite{Gelmini:2013awa}
\bibitem{Gelmini:2013awa}
  G.~B.~Gelmini, J.~H.~Huh and T.~Rehagen,
  %``Asymmetric dark matter annihilation as a test of non-standard cosmologies,''
  arXiv:1304.3679 [hep-ph].  %%CITATION = ARXIV:1304.3679;%%  %1 citations
  %%counted in INSPIRE as of 01 Aug 2013

\bibitem{Wang:2015gua}
  S.~Z.~Wang, H.~Iminniyaz and M.~Mamat,
  %``Relic Abundance of Asymmetric Dark Matter in Scalar--Tensor Model,''
  arXiv:1503.06519 [hep-ph].
  %%CITATION = ARXIV:1503.06519;%%



%\cite{Meehan:2014zsa}
\bibitem{Meehan:2014zsa}
  M.~T.~Meehan and I.~B.~Whittingham,
  %``Asymmetric dark matter in braneworld cosmology,''
  JCAP {\bf 1406}, 018 (2014)
  [arXiv:1403.6934 [astro-ph.CO]].
  %%CITATION = ARXIV:1403.6934;%%
  %3 citations counted in INSPIRE as of 26 Mar 2015



\bibitem{Lahav:2014vza}
  O.~Lahav and A.~R.~Liddle,
  %``The Cosmological Parameters 2014,''
  arXiv:1401.1389 [astro-ph.CO].
  %%CITATION = ARXIV:1401.1389;%%
  %13 citations counted in INSPIRE as of 24 Apr 2015
%arXiv:1001.4538 [astro-ph.CO];


\bibitem{Ackermann:2011wa}
 M.~Ackermann {\it et al.}  [Fermi-LAT Collaboration],
 Phys.\ Rev.\ Lett.\  {\bf 107}, 241302 (2011)  [arXiv:1108.3546 [astro-ph.HE]].
  %``Constraining Dark Matter Models from a Combined Analysis of Milky Way Satel%lites with the Fermi Large Area Telescope,''


\bibitem{standard-cos}
R.~J.~Scherrer and M.~S.~Turner, Phys. Rev. D {\bf 33}, 1585 (1986),
Erratum-ibid. D {\bf 34}, 3263 (1986).

\bibitem{wmap}
  G.~Hinshaw {\it et al.}  [WMAP Collaboration],
  %``Nine-Year Wilkinson Microwave Anisotropy Probe (WMAP) Observations: Cosmological Parameter Results,''
  Astrophys.\ J.\ Suppl.\  {\bf 208}, 19 (2013);
%  [arXiv:1212.5226 [astro-ph.CO]].
  %%CITATION = ARXIV:1212.5226;%%
  %1412 citations counted in INSPIRE as of 10 Mar 2015
%\bibitem{Bennett:2012zja}
  C.~L.~Bennett {\it et al.}  [WMAP Collaboration],
  %``Nine-Year Wilkinson Microwave Anisotropy Probe (WMAP) Observations: Final Maps and Results,''
  Astrophys.\ J.\ Suppl.\  {\bf 208}, 20 (2013).
%  [arXiv:1212.5225 [astro-ph.CO]].
  %%CITATION = ARXIV:1212.5225;%%
  %640 citations counted in INSPIRE as of 11 mar 2015







\end{thebibliography}
\end{document}